\documentclass[10pt,twoside]{article}
\usepackage[paperheight=241.2mm, paperwidth=168.7mm,
top=20mm,textheight=189.8mm,left=2cm,textwidth=127.4mm,
includefoot=false,footskip=8mm,ignorehead]{geometry}
\usepackage[utf8]{inputenc}
\usepackage{ngerman}
\usepackage{fancyhdr}
\usepackage{garamond}
\usepackage[T1]{fontenc}
\usepackage{graphicx}
\usepackage{wrapfig}
\usepackage{fmtcount}
\usepackage{rotating}
\usepackage[table]{xcolor}
\usepackage{tabularx}
\usepackage[footnotesize]{caption}
\usepackage{indentfirst}
\usepackage{ziffer}

\makeatletter
\renewcommand\section{\@startsection
   {section}{1}{0mm}
   {1.5\baselineskip}
   {0.2\baselineskip}
   {\normalsize\centering\bfseries}
   }
\makeatother 

\makeatletter
\renewcommand*{\@seccntformat}[1]{\csname the#1\endcsname\hspace{0.5em}}
\makeatother

\newcommand{\dg}{{\raisebox{0.4em}{\tiny $\circ$}\ }}
\newcommand{\dm}{{\raisebox{-0.0em}{\small \grq}\ }}
\newcommand{\ds}{{\raisebox{-0.0em}{\small \grqq}\ }}
\newcommand{\am}{\raisebox{0.2em}{\footnotesize m} }
\newcommand{\as}{\raisebox{0.2em}{\footnotesize s} }
\newcolumntype{C}[1]{>{\hsize=#1\hsize\centering\arraybackslash}X}

\begin{document}
\setcounter{page}{69}
\garamond
\renewcommand{\baselinestretch}{0.95}
\small

\begin{center}
\large{\textbf{Die weltweit erste Messung einer Lotabweichung in der Länge}} 

\vspace{2\parskip}
\small{Andreas Schrimpf (Marburg)}
\end{center}
\vspace{-\parskip}

{\footnotesize 
\baselineskip9.5pt
\noindent
\textbf{Abstract.}
During the summer of $1837$ Christian Ludwig Gerling, a former student of Carl Friedrich Gauss, 
organized the world wide first determination of a vertical deflection in longitude. 
From a mobile observatory at the Frauenberg close to Marburg (Hessen) he measured the 
astronomical longitude difference between the observatory of C.F.\ Gauss at Göttingen and 
the observatory of F.G.B.\ Nicolai at Mannheim. By comparing these astronomical results with 
the geodetic determined longitude differences, which he just had measured as a part of the 
triangulation of Kurhessen, he was able to extract a combined value of the vertical deflection 
of Göttingen and Mannheim. His results are in very good agreement with modern vertical deflection 
data.

\noindent
\textbf{Zusammenfassung.}
Im Sommer $1837$ führte Christian Ludwig Gerling, Schüler von Carl Friedrich Gauß, die weltweit 
erste Bestimmung einer Lotabweichung in der Länge durch. Vom Frauenberg bei Marburg aus ermittelte 
er den astronomischen Längenunterschied der Sternwarte von C.F.\ Gauß in Göttingen zu seiner 
Messstation am Frauenberg und zur Sternwarte von F.G.B.\ Nicolai in Mannheim. Durch Vergleich 
mit den von ihm im Rahmen der Kurhessischen Triangulierung gemessenen geodätischen 
Längendifferenzen erhielt er die Lotabweichungen in der Länge. Möglich wurde dieses Ergebnis 
nur durch Gerlings höchst sorgfältige Arbeitsweise, durch ein Ausnutzen der maximalen Möglichkeiten 
der ihm zur Verfügung stehenden Messgeräte. Seine Ergebnisse stimmen gut mit den heutigen Daten der 
Lotabweichungen überein.

}

\section{Einleitung}
Die Erforschung und Diskussion der Gestalt der Erde hat in der Geschichte der Menschheit sehr 
deutliche Spuren hinterlassen. Ohne sehr präzise Messinstrumente spricht die Erfahrung der 
Menschen für eine flache, ebene Welt. Jedoch sind bereits in der Antike die Hinweise auf eine 
kugelförmige Form der Erde weitgehend anerkannt. Interpretiert man die Beobachtung der täglichen 
Bewegung der Gestirne am Himmelsfirmament als eine scheinbare Bewegung aufgrund der 
Eigendrehung der Erde, dann ermöglicht die Rotationsachse der Erde die Definition einer 
Himmelssphäre, des äquatorialen (sphärischen) Koordinatensystems, in dem alle Gestirne mit 
zwei Winkeln platziert werden können. Durch eine Vermessung der Höhen der Sterne (z.B. der Polhöhe) 
kann man die geographische Breite und durch eine Bestimmung der Durchgangszeit der Sterne 
im Zenitbogen die geographische Länge auf der Oberfläche der angenommenen Erdkugel berechnen. 
Vergleicht man nun diese aus astronomischen Beobachtungen erhaltenen Koordinaten mit solchen, 
die durch eine Landvermessung der Erdoberfläche selber bestimmt wurden, so ergibt sich ein 
differenziertes Bild der Gestalt der Erde: zunächst ist unser Heimatplanet im Mittel durch ein 
an den Polen abgeplattetes Rotationsellipsoid geometrisch beschreibbar, der Abstand von Pol zu 
Pol ist um etwa $43$ km geringer als der Durchmesser am Äquator. Weiterhin findet man vertikale 
Abweichungen der physikalischen Erdfigur (dem Geoid) von diesem Rotationsellipsoid im Bereich
von etwa $\pm 100$~m. Das Geoid entspricht der Niveaufläche des mittleren globalen Meeresspiegels, 
die man sich unter den Kontinenten fortgesetzt vorstellen kann. Den Winkelunterschied zwischen 
der auf dem Geoid senkrecht stehenden physikalisch definierten Lotlinie von der geometrisch 
definierten Ellipsoidnormalen nennt man Lotabweichung; diese lässt sich sehr präzise durch eine 
Vermessung der tatsächlich im Zenit stehenden Sterne im Vergleich zu den auf dem Referenzellipsoid 
erwarteten Zenitsternen bestimmen.
Ist schon die Abplattung der Erde ein kleiner Effekt im Bereich nur einiger Promille, so sind die 
Abweichungen des Geoids vom Referenzellipsoid nochmals um etwa $3$ Größenordnungen kleiner. Die 
Entdeckung und Vermessung der Lotabweichungen ist daher sehr durch die in der jeweiligen Zeit 
erreichbare Empfindlichkeit der Messmethoden geprägt.

\section{Die Erde als Rotationsellipsoid}
\indent
Die Vermessung der Gestalt der Erde stellte für viele Jahrhunderte eine große Herausforderung dar. 
In der Antike war es Eratosthenes (um $240$ v.Chr.), der die Kugelgestalt der Erde zu Grunde legte 
und ihren Radius aus einer gleichzeitigen Messung des Zenitwinkels der Sonne an zwei 
verschiedenen Orten und der Weglänge, d.h. der Bogenlänge auf einem Meridian zwischen 
diesen beiden Orten abschätzte. Damit legte Eratosthenes die Grundlage der so genannten 
,,Gradmessung'': Man ermittelt den Abstand zweier Orte, die zueinander genau in Nord-Südrichtung
gelegen sind, auf der Erdoberfläche und vergleicht dies mit dem aus den zugehörigen Polhöhen von 
Gestirnen vermessenen Himmelsbogen. Man erhält dadurch eine Information über die geometrische 
Gestalt der Erde längs eines Meridians, konkret bezüglich der Streckenlänge eines geographischen 
Breitenunterschiedes. Eratosthenes selber führte die astronomische Breitenmessung durch und 
übernahm die Angaben zur Länge des Bogens auf der Erde vermutlich aus Angaben von Schrittzählern 
des ägyptischen Katasters (Torge $2009$). Im Vergleich zu heutigen Daten ermittelte er den Radius 
mit einer Genauigkeit von etwa $10$ \%. Eine etwas genauere Messung beauftragte im frühen
Mittelalter der Bagdader Kalif Al-Ma’mun (um $820$). Er ließ einen $2$ Grad langen Bogen möglichst 
exakt bestimmen und erhielt den Erdradius mit einem Fehler von nur etwa $1$--$2$ \%.

Im $17$. Jahrhundert begannen die Wissenschaftler sich im Zusammenhang mit der Entwicklung des
Konzepts der Gravitation Gedanken um den inneren Aufbau der Erde, deren mittleres spezifisches 
Gewicht und ihrer genauen Form zu machen. Mit einem Fadenpendel, dessen Schwingungsperiode 
proportional zur Wurzel aus dem Verhältnis von Länge und Gravitationsfeldstärke (Erdbeschleunigung) 
ist, wurden erste Anzeichen systematischer Abweichungen im Schwerefeld der Erde an Orten 
unterschiedlicher geographischer Breite festgestellt. Die Interpretation dieser Hinweise lieferte 
Isaac Newton: ein homogener, flüssiger und rotierender Erdkörper sollte die Gestalt eines 
Rotationsellipsoides annehmen, also am Äquator und an den Polen
unterschiedliche Krümmungen aufweisen. Zur Bestätigung dieser Vermutung veranlasste die französische Akademie der 
Wissenschaften Gradmessungen in hoher und niedriger geographischen Breite, durchgeführt $1736$/$37$ 
von Pierre-Louis Moreau de Maupertuis und Alexis-Claude Clairaut in Lappland und von $1735$--$1744$ 
von Pierre Bouguer, Charles-Marie de la Condamine und Louis Godin in Peru (Torge $2009$). Die 
Ergebnisse dieser Expeditionen ergaben eine an den Polen leicht abgeflachte ,,Kugel'' mit einer 
Abplattung von etwa $1$/$300$ –-- das Rotationsellipsoid als geometrische Form der Erde war geboren!

Wohl war den Wissenschaftlern bewusst, dass Gebirge und Meerestiefen nicht durch einen 
rotationsymmetrischen Körper beschrieben werden konnten, aber im Mittel, so die Idee, 
sollte die Erde, die sich ja mit sehr hoher Präzision gleichmäßig dreht, einem geometrischen 
Rotationskörper sehr nahe kommen. Mit dieser Idee und einer verbesserten Messtechnik 
ausgestattet führten Pierre Méchain und Jean- Baptiste Joseph Delambre wiederum im Auftrag 
der französischen Akademie der Wissenschaften von $1792$ bis $1798$ eine große Gradmessung durch 
(Alder $2005$). Ihr Ziel war es, die Streckenlänge für einen etwa $10$\dg  großen Meridianbogen der 
Erde durch Dreiecksmessungen möglichst exakt zu bestimmen und aus dem Vergleich mit den 
astronomisch ermittelten Polhöhen der Enden dieses Bogens und deren daraus abgeleiteten 
Breiten nicht nur die Größe der Erde mit bisher ungekannter Präzision zu bestimmen, sondern 
daraus auch ein Normmaß, das Meter, als Maßstab abzuleiten, welches überall auf der Erde durch 
eine vergleichbare Messung ebenso bestimmt werden kann.

So erfolgreich diese Messung in ihrer Methodik, in der Anwendung neuer Messgeräte war, so 
offenbarte sie doch ein überraschendes Ergebnis: der durch Paris verlaufende vermessene 
Meridianboden zeigte eine etwa doppelt so große Krümmung, eine Abplattung von
$1$/$150$ (Laplace $1799$, S.\ $138$ ff), wie der aus bisherigen Messungen
ermittelte Wert. So kombinierte Delambre die Daten der französischen Gradmessung
mit denen der Gradmessung in Peru und verwendete für die Berechnung des Meters eine Abplattung von $1$/$334$ (Torge $2009$). Méchain und 
Delambre und in den folgenden Jahren weitere Wissenschaftler fanden sich zunehmend mit der 
Erkenntnis ab, dass jeder Meridian der Erde eine eigene Krümmung aufweise, dass die genaue 
Gestalt der Erde sich nicht durch einen rotationssymmetrischen Körper beschreiben ließe. 
Dennoch konzentrierte sich die Wissenschaft im auslaufenden $18$. und beginnenden $19$. Jahrhundert 
zunächst erst einmal auf eine möglichst gute Bestimmung des Referenzellipsoids als 
Beschreibung der mittleren Gestalt der Erde. Der von Henrik Johan Walbeck aus 5 Gradmessungen 
bestimmte Wert von $1$/$302,78$ (Gauß $1828$) wurde u.a.\ sowohl von Gauß als auch von Gerling 
für deren Ausgleichsrechnungen verwendet. Mit hoher Sorgfalt bestimmte $1841$ Friedrich Wilhelm 
Bessel aus $10$ verschiedenen Gradmessungen und einer Korrektur der Berechnungen der französischen 
Meridianmessung die mittlere Abplattung zu $1$/$299,1528$ (Bessel $1837$ und Bessel $1841$) und 
seit $1979$ stellt das GRS$80$-Ellipsoid mit $1$/$298,257222101$ die allgemein empfohlene beste 
Beschreibung des globalen Referenzellipsoids dar.

\section{Christian Ludwig Gerling}

\begin{wrapfigure}{r}{0.46\textwidth}
\vspace{-22pt}
\begin{center}
\includegraphics[angle=0,width=0.44\textwidth]{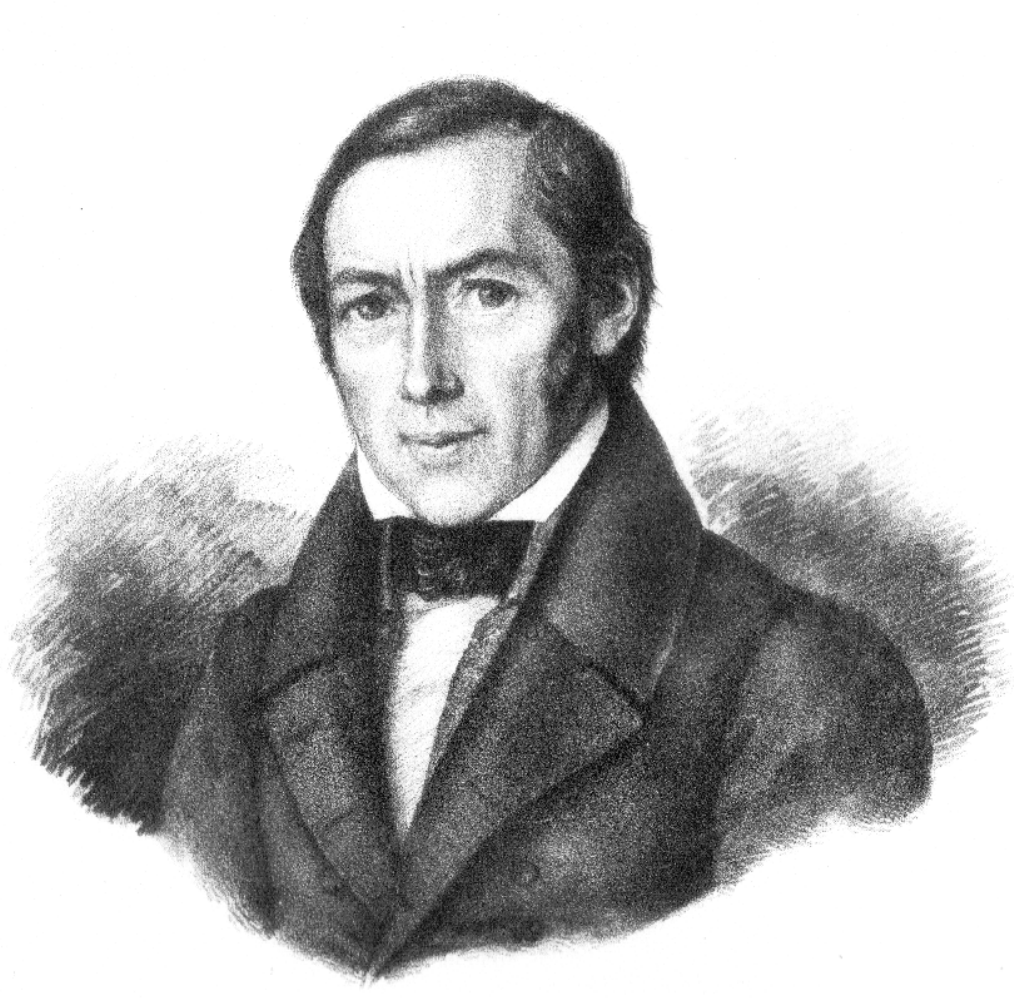}
\vspace{-8pt}
\caption{Christian Ludwig Gerling ($1788$--$1864$)} 
\end{center}
\vspace{-20pt}
\end{wrapfigure}

Christian Ludwig Gerling (Abb.\ $1$) wurde $1788$ in Hamburg als Sohn eines Pfarrers geboren. Im Gymnasium lernt er 
seinen langjährigen Freund Johann Franz Encke, später Direktor der Berliner Sternwarte, kennen. 
Nach dem Abitur studierte Gerling zunächst in Helmstedt Theologie, nahm dann aber nach der Auflösung 
dieser Universität $1810$ sein Studium in Göttingen auf. Dort begegnete er Carl Friedrich Gauß und 
wechselte 
unter dessen Einfluss zu den Fächern Mathematik, Astronomie, Physik und Chemie. Er 
arbeitete unter Leitung 
von Gauß und Karl Ludwig Harding in der Göttinger Sternwarte, besuchte im Sommer $1811$ die 
Sternwarten in 
Gotha, Halle und Leipzig und beendete $1812$ seine Promotion über die Berechnung des Verlaufs der 
Sonnenfinsternis von $1820$ durch Europa.

Während Gerlings Zeit in Göttingen traf er seinen alten 
Schulfreund Encke wieder, und begegnete u.a.\ auch Friedrich Gottfried Bernhard
Nicolai ($1793$–-$1846$), der später die Leitung der Sternwarte in Mannheim übernahm. Gerling 
nahm nach 
seiner Promotion eine Stelle als Lehrer eines Gymnasiums in Kassel an und erhielt schließlich 
$1817$ einen 
Ruf als ordentlicher Professor der Mathematik, Physik und Astronomie an die Philipps-Universität Marburg. 
Er bliebt der Universität Marburg bis zu seinem Tode im Jahre $1864$ treu (Madelung $1996$).
Gerlings wissenschaftliches Werk ist von zwei Dingen geprägt: in seiner frühen Zeit in Marburg beschäftigte 
ihn sehr stark die Triangulation Kurhessens. Parallel dazu bewegten Gerling auch weiterhin astronomische 
Themen. Er konnte allerdings erst $1841$ ein eigenes Institutsgebäude mit einer 
Sternwarte\footnote{Die Sternwarte prägt auch heute noch weit sichtbar das Gesicht des Fachbereichs
 Physik der Philipps-Universität Marburg. 
Siehe Webseite http://www.parallaxe-sternzeit.de des Fördervereins, der sich um Bewahrung und 
Nutzung des Erbes Gerlings bemüht (Schrimpf/Lipphardt/Heckmann $2010$).} in Marburg in 
Betrieb nehmen und widmete sich daher vor allem in der späteren Phase seines Schaffens etwas stärker 
astronomischen Fragen der Zeit, wie z.B. der Beobachtung von Planeten und Asteroiden, der genaueren 
Vermessung der astronomischen Einheit und der Verbesserung der Sternkarten. 

Mit Carl Friedrich Gauß 
verband Gerling eine lebenslange Freundschaft, zunächst mehr als seinem Lehrer und Mentor, in den 
späteren Jahren aber beiderseitig als Freunde und Ratgeber. Ein intensiver Briefverkehr zwischen beiden 
enthält neben wissenschaftlichen Details, die in Publikationen nicht zu finden sind, auch ein beredtes 
Zeugnis der Beziehung der beiden (Schäfer $1927$, Gerardy $1964$). 
Gerling lernte bei Gauß nicht nur den sehr sorgfältigen Umgang mit Messgeräten, die Beachtung und 
Berücksichtigung aller Messfehler, sondern auch die mathematischen Methoden, die zur Auswertung 
geodätischer und astronomischer Messungen notwendig sind. Gerling ist nicht nur der Herausgeber eines 
Lehrbuches der ebenen und sphärischen Trigonometrie (in mehreren Auflagen, die letzte posthum $1865$), 
sondern auch der Vermittler vieler von Gauß entwickelter Methoden, so u.a.\ der Methode der kleinsten 
Fehlerquadrate in der Anwendung für geodätische Aufgaben (Gerling $1843$). So schreibt Johann 
Jacob Baeyer ($1864$) als Nachruf auf Gerling: 

„\textit{Die mitteleuropäische Gradmessung hat in ihm einen 
Geodäten ersten Ranges mit reichen Erfahrungen verloren. Er war der einzige noch lebende Mitarbeiter 
von Gauss an der Hannöverschen Gradmessung, und mit der Methode seines grossen Lehrers, der selbst 
nichts darüber hinterlassen, vollständig vertraut, so dass er manchen Aufschluss hätte geben können 
über Fragen, die nun vielleicht für immer im Dunkeln gehüllt bleiben.}'' 

\section{Die kurhessische Triangulierung}
Im Frühjahr $1821$ bat Kurfürst Wilhelm II von Hessen Gerling um ein Gutachten über eine 
Vermessung zur Grundlage einer topographischen Karte Kurhessens. Nach einer ersten
geowissenschaftlichen Erkundung im Herbst $1821$ und Frühjahr $1822$ erhielt Gerling $1822$ 
dann den Auftrag zu Erstellung eines Hauptdreiecknetzes von Kurhessen (Gerling $1839$, 
Reinhertz $1901$). Er führte die Vermessungen in zwei Perioden von $1822$-$1824$ und $1835$-$1837$ 
durch und schloss im Norden an die Hannoversche Gradmessung seines Lehrers an; dabei bezog er 
u.a.\ auch das große Gaußsche Dreieck Brocken-Hohehagen-Inselsberg mit ein. Im Süden 
verband Gerling sein Netz mit einigen Punkten der alten bayerischen und der alten 
großherzoglich-hessischen Landestriangulation. 

Für seine Messungen verwendete Gerling einen $12$-zölligen Repetitionstheodoliten von 
Rei-chenbach-Ertel und mindestens ein $10$-zölliges Universalinstrument von Breithaupt. Als 
Längenreferenz bei den lokalen Zentrierungsmessungen diente eine Kopie der Toise von Peru 
(auf der die damalige Meter-Definition beruhte), die er sich vor der zweiten Messperiode von 
Fortin aus Paris beschafft hatte\footnote{Das Breithauptsche Universalinstrument und die 
Toise befinden sich heute in der Physikalischen Sammlung der Philipps-
Universität Marburg.}. Gerlings Triangulation umfasst $24$ Punkte I.\ Klasse und $17$ 
Punkte II.\ Klasse. Als Basislänge nutzte Gerling die von Gauß aus Heinrich Christian 
Schumachers Messungen in Holstein abgeleitete Linie ,,Sternwarte-Meridianzeiche'' in Göttingen. 
Gerling legte seiner Haupttriangulation das Ellipsoid von Walbeck zu Grunde und orientierte sein 
Netz mit den von Gauß in Göttingen durchgeführten Polhöhen- und Azimutbestimmungen. Er berechnete 
alle $24$ Punkte I.\ Klasse in einem Guss durch Netzausgleichung\footnote{Die Punkte II. Klasse 
konnten wegen des hohen Aufwandes nicht in die Ausgleichsrechnung eingeschlossen werden und 
wurden nachträglich hinzugefügt. Auch wurden sie mit geringerer Genauigkeit vermessen.}, 
ein enormer rechnerischer Aufwand, 
für den Gauß ihm großen Respekt zollte (Brief Nr.\ $290$ von Gauß, Schäfer $1927$). Gerling berechnet 
den mittleren Fehler seiner Richtungen zu $\pm 0,88$\ds; eine kritische Nachrechnung von Börsch im 
Zuge der mitteleuropäischen Gradmessung ergibt einen mittleren Winkelfehler von $\pm 0,946$\ds 
(Baeyer $1866$). Gerlings Arbeiten stellen für Hessen die erstmalige Berechnung eines 
Triangulationsnetzes auf einem Rotationsellipsoid dar (Heckmann $2012$).

Fast $175$ Jahre nach Gerlings Pionierarbeit hat das Hessische Landesamt für Bodenmanagement und 
Geoinformationen (HLBG)eine Bestandsaufnahme der im amtlichen Festpunktnachweis enthaltenen 
kurhessischen Hauptdreieckspunkte Gerlings durchgeführt (Heckmann $2012$). Da Gerling seine 
Messpunkte zum Teil mit tonnenschweren Steinen markierte, sind diese heute noch gut erhalten. 
Das Ergebnis ist erstaunlich: $14$ Punkte I.\ Klasse und $6$ Punkte II.\ Klasse sind noch an originaler 
Lage identitätssicher erhalten. Gezielte Nachforschungen ermöglichten dann sogar die 
Identifizierung von $2$ weiteren noch vorhandenen Punkten I.\ Klasse, einem erhaltenen 
Punkt II.\ Klasse sowie der exakten Rekonstruktion eines dritten Punktes I.\ Klasse. 
Anschließend ergab ein Vergleich der Gerlingschen kurhessischen Positionsangaben mit den 
hochgenauen Koordinaten des heutigen geodätischen Bezugssystems über ganz Hessen betrachtet 
eine Lagequalität von meist besser als $20$~cm und nur in seltenen Ausnahmen (an den Netzrändern) 
von schlechter als $30$~cm (Heckmann $2012$).

\captionsetup{justification=centering}
\begin{figure}[t]
\begin{minipage}[b]{0.38\textwidth}
\caption{Netzbild der Kurhessischen\\ Triangulierung von $1822$ -– $1837$\\ (Gerling $1839$)}
\end{minipage}
\hspace{2em}
\begin{minipage}[b]{0.55\textwidth}
\includegraphics[angle=0,width=\textwidth]{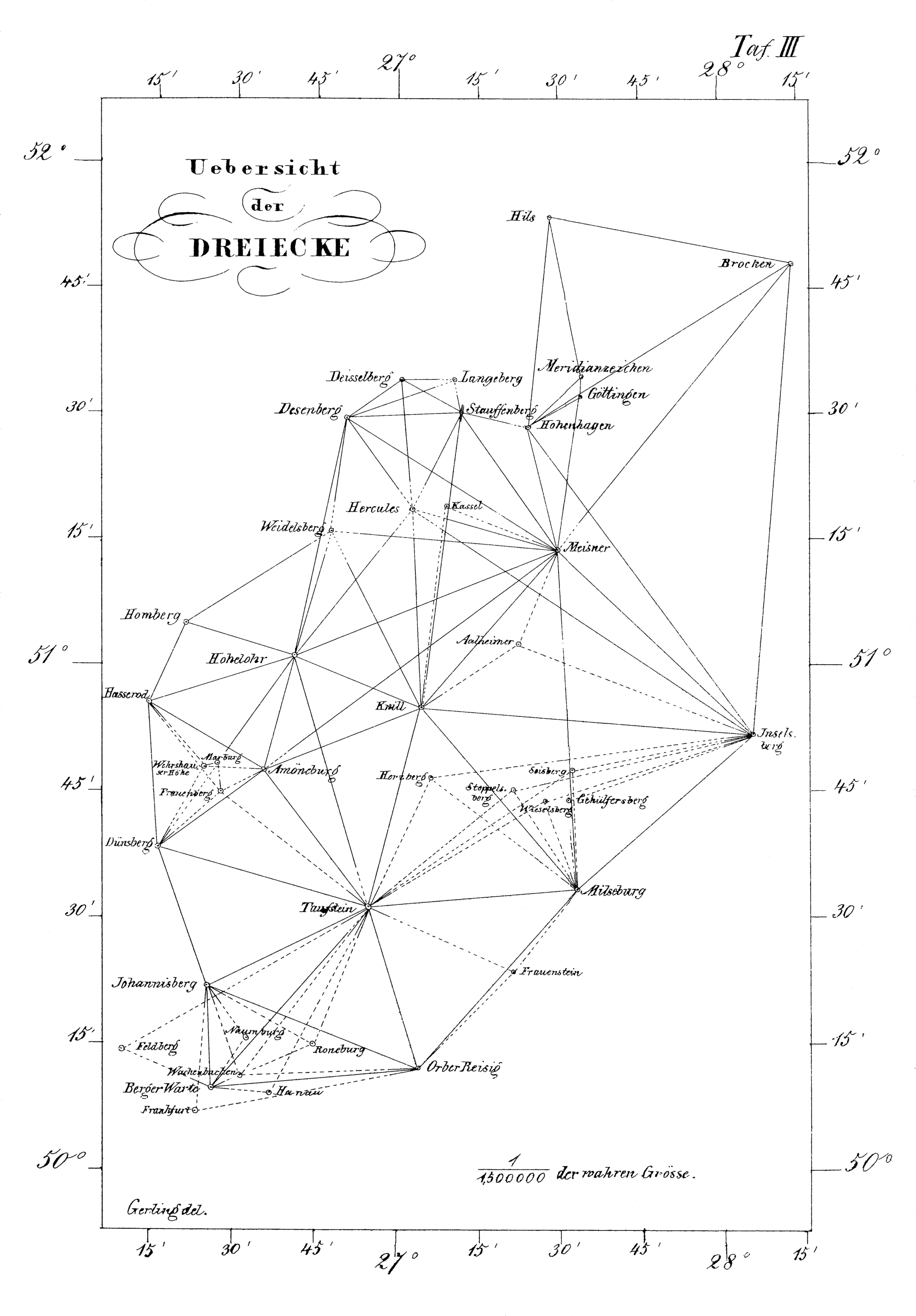}
\end{minipage}
\end{figure}

\section{Die Anfänge der Lotabweichungsmessungen}
Die Gradmessungen im $18$.\ und beginnenden $19$.\ Jahrhundert offenbarten in immer stärkerem Maße 
lokale Abweichungen der Erdfigur von einem Rotationsellipsoid. Wegen des starken Einflusses der 
Abplattung auf die Bestimmung des Mondortes schlägt z.B.\ Johann Georg von Soldner vor, für die 
Beobachtung eine Sternwarte am Äquator in Afrika zu errichten und den zugehörigen Meridian 
genau zu vermessen (Torge $2009$). Pierre-Simon Laplace und von Soldner regen auch die 
Einführung einer breiten- bzw. längenabhängigen Abplattung an (Torge $2009$).
Jedoch waren die bisherigen Ergebnisse eher zusätzliche Erkenntnisse aus Messungen, die 
eigentlich der genaueren Bestimmung der rotationssymmetrischen Gestalt der Erde dienen sollten.

Im Zuge der Diskussion um die Ableitung eines einheitlichen Maßstabs aus dem Umfang der Erde 
wurde als alternative Messmethode das so genannte Sekundenpendel weiter entwickelt und 
perfektioniert. Da die Schwingungsdauer eines mathematischen Pendels nur von dessen Länge und 
der Schwerebeschleunigung abhängt, kann man damit bei bekannter Beschleunigung ein Längennormmaß 
konstruieren oder aber auch bei bekannter Länge des Pendels die Schwerebeschleunigung absolut 
messen. Man hatte bereits im $18$.\ Jahrhundert begonnen, parallel zu Gradmessungen auch die 
lokale Schwerefeldstärke mit solchen Pendeln zu vermessen. Ist die Erde kein rotationssymmetrischer 
Körper, dann müssen sich die Höhenabweichungen auch in Differenzen der Schwerefeldstärke äußern.
Dies ist die Grundidee der Gravimetrie, die zweite wichtige Methode neben der 
Astrogeodäsie (=Vergleich astronomischer und geodätischer Ortsbestimmungen) zur genauen 
Vermessung der Form Erde. $1818$ beschrieb Henry Kater zum ersten Mal die Nutzung eines 
verbesserten Sekundenpendels, eines Reversionspendels (Kater $1818$). Eine begrenzte Anzahl von 
Messungen wurde ab etwa $1860$ ausgeführt, als endlich transportable Geräte entwickelt waren. 
Jedoch erst im $20$.\ Jahrhundert verhalfen so genannte Freifall-Gravimeter der Schweremessung 
zum wirklich großen Durchbruch.

Vermutlich die erste bewusste Bestimmung einer lokalen Anomalie der Oberfläche der Erde führte 
C.F.\ Gauß $1827$ durch und erläutert diese in seinem Bericht über die Bestimmung des 
Breitenunterschieds zwischen den Sternwarten Göttingen und Altona (Gauß $1828$). Gauß bestimmte 
an beiden Standorten mit einem Ramsdenschen Zenitsektor die Zenitdistanzen einiger Sterne. 
Dann verglich er die daraus gewonnenen astronomischen bestimmten Breiten mit den bei seiner 
Gradmessung im Königreich Hannover von $1821$–$1824$ ermittelten geodätischen Breiten. Es ergab 
sich ein um $5,52$\ds geringerer astronomischer Breitenunterschied. Unter Ausnutzung einer von 
Franz Xaver von Zach auf dem Brocken gemessenen astronomischen Polhöhe findet er zwischen 
Göttingen und dem Brocken einen um $10$-$11$\ds größeren astronomischen Breitenunterschied und 
daraus zwischen Altona und dem Brocken eine Breitendifferenz von $16$\ds\hspace{-0.5ex}! Gauß stellt in seiner 
Publikation fest, dass solch ein Unterschied nichts Außergewöhnliches sei, ja dass man dies, 
wenn nur die Genauigkeit der Messmethoden um ein bis zwei Größenordnungen höher sei, sicher 
an jedem Punkt der Erde feststellen könne. Jedoch verweist er darauf, dass erst 
,,\textit{die künftigen Jahrhunderte die mathematische Kenntniss der Erdfigur
sehr viel werden weiter bringen können}'' (Gauß $1828$; näheres hierzu siehe z.B.\ 
Wittmann et al.\ $2010$).

Mit dieser nicht gerade ermutigenden Anmerkung lag Gauß sehr richtig. Auffallend ist an all den 
bisher aufgeführten astro-geodätischen Messungen, dass diese immer nur den astronomisch und 
geodätisch ermittelten Breitenunterschied im Fokus hatten und jeweils auf demselben 
Meridianbogen vermessen wurden. Um die unregelmäßige Form der Erde zu beschreiben, führte 
man den Begriff der Lotabweichung ein: das gemessene Lot einer Beobachtungsstation differiert 
um die Lotabweichung von der Richtung des Lots auf dem Referenzellipsoid 
(Definition nach Helmert, Torge $2003$). Dabei wird die Abweichung auf der Referenzfläche im 
Allgemeinen durch zwei Koordinatenrichtungen beschrieben, zum Beispiel (und am sinnvollsten) 
durch eine Lotabweichung in der Breite (d.h.\ in Süd-Nord-Richtung) und eine in der Länge (d.h.\ 
in West-Ost-Richtung). Heute gibt man die Lotabweichungskomponenten 
$\xi$ und $\eta$ wie folgt an (Torge $2003$):
\[ \xi = \phi - \varphi \mbox{\hspace*{1cm}}  \eta = (\Lambda - \lambda)
\cos\phi
\]

\noindent
Dabei bedeuten 	\hspace*{2mm} $\phi$   die astronomische Breite,\\
\hspace*{2.3cm} $\varphi$	  die geodätische Breite,\\
\hspace*{2.3cm} $\Lambda$	 die astronomische Länge und\\
\hspace*{2.3cm} $\lambda$	 die geodätische Länge.

Während man für eine Breitenmessung die Zenithöhen von Sternen (bzw.\ die Polhöhe des 
Beobachtungsortes) feststellt, muss bei einer Längenbestimmung die Durchgangszeit von Sternen 
im Meridian notiert werden. Längenbestimmungen sind also Zeitmessungen. Sind schon 
Breitenmessungen mit der erforderten Präzision sehr aufwendig, so gilt dies umso mehr für 
Längenmessungen. Um eine Längenmessung mit einer Genauigkeit von etwa einer Zehntel 
Bogensekunde ($0,1$\ds\hspace{-0.5ex}) durchzuführen, bedarf es einer Genauigkeit in der 
Zeitmessung von $0,1$/$15$ sec = $0,006$ Sekunden! Mit dieser Genauigkeit muss eine 
Zeitdifferenz an verschiedenen Orten der Erde bestimmt werden, um eine Präzision 
von $0,1$\ds in der Lotabweichung der Länge zu erreichen, keine sehr einfache Aufgabe 
im $19$.\ Jahrhundert!

Im Jahr $1824$ unternahm Nicolai eine Längenbestimmung in der Gegend von Mannheim, die man als 
Vorläuferarbeit zu Gerlings Messungen werten kann (Nicolai $1825$). Als Teil einer von der 
französischen Regierung vorgeschlagenen Vermessung eines Längenbogens ermittelte Nicolai 
gemeinsam mit dem französischen Ingenieur-Obristen Henry sowie mit Johann Gottlieb 
Friedrich von Bohnenberger und Friedrich Magnus Schwerd die astronomischen Längendifferenzen 
von Straßburg über Tübingen und Speier bis nach Mannheim. Mittels Pulversignalen wurden die 
Uhren der vier Sternwarten synchronisiert und die jeweilige lokale Sternzeit anhand des 
Durchgangs einiger Bessel‘scher Fundamentalsterne gemessen. Die Differenz der Sternzeiten 
entsprach dann den gesuchten Längenunterschieden. Ein Vergleich ergab
,,\textit{eine vortreffliche Übereinstimmung}'' (so Nicolai) der geodätischen und
astronomischen Daten für die Differenz Mannheim-Straßburg (geodätisch $2$\dm $54,05$\ds\hspace{-0.5ex}), einen 
Unterschied von $0,35$\ds für die Längendifferenz 
von Mannheim zu Tübingen (geodätisch $2$\dm $21,91$\ds\hspace{-0.5ex}) und eine Abweichung von $0,16$\ds für die 
Längendifferenz von Mannheim zu Speier (geodätisch $4,90$\ds\hspace{-0.5ex}). Nicolai verstand die astronomischen 
Messungen als Bestätigung der geodätischen Daten und erwartete ganz offensichtlich auch keinen 
Unterschied!

\section{Gerlings Längendifferenz-Messungen am Frauenberg im Sommer 1837}
Zum Abschluss der kurhessischen Triangulierung beabsichtigte Gerling --- genau
wie auch Nicolai zuvor, eine Kontrollmessung durchzuführen, eine astronomische
Längenbestimmung über das von ihm geschaffene Dreiecksnetz (Gerling $1838$). Da
Sternwarten mit größeren standfesten Geräten präzisere Ergebnisse ermöglichen,
wählte er die Sternwarten Göttingen im Nordosten und Mannheim, etwa auf der gleichen Länge wie der am südwestlichen Netzrand befindliche Feldberg gelegen, aus. 
Bei diesen Messungen halfen in Göttingen Gauß und dessen Assistent Carl Benjamin Goldschmidt, in 
Mannheim unterstützte ihn Nicolai. Gerling selber verfügte $1837$ noch nicht über eine Sternwarte 
und wählte daher als Messpunkt in der Nähe von Marburg den Frauenberg, eine kleine Erhöhung etwa 
$6$~km südöstlich von Marburg gelegen. 
Am Frauenberg hatte Gerling im Rahmen seiner kurhessischen Triangulierung mithilfe eines 
Stangensignales einen Dreieckspunkt II.\ Klasse auf der dortigen Ruine festgelegt. Für die nun 
geplanten weitergehenden Messungen stellte Gerling zunächst einen schweren Steinpfeiler auf, 
den er als Standort für den $10$-Zoll Theodolit nutzen wollte. Ein weiterer sehr wichtiger Teil 
seiner Messstation war ein Box-Chronometer von Kessels, eine Präzisionsuhr, die er kurz zuvor 
angeschafft hatte und die ihm später als Sternwartenuhr diente.
Die Aufgabe bestand darin, die Uhren der drei Sternwarten höchst präzise zu synchronisieren und 
dann an jedem Standort eine Sternzeitbestimmung vorzunehmen, um daraus den astronomischen 
Längenunterschied zu ermitteln. Gerling wählte die Zeit zwischen dem $24$.\ August und dem $9$.\ 
September $1837$ für das geplante Messprogramm aus.

Die Sternzeitmessungen waren eine Routinetätigkeit der damaligen Sternwarten. Die aus 
Durchgangsmessungen ermittelten lokalen Sternzeiten von Göttingen und Mannheim erhielt Gerling 
von Gauß und Nicolai. Der Steinpfeiler am Frauenberg stellte sich als unsicheres Fundament 
für den Theodolit heraus, da er zu frisch gesetzt war und der Boden noch etwas nachgab. So hat 
Gerling die lokale Zeit anhand von korrespondierenden Sonnenhöhen vermessen; diese Ergebnisse 
waren genauer als die Durchgangsmessungen mit dem Theodoliten auf dem noch unstabilen Steinpfeiler.

Die wirkliche Herausforderung bestand in der Synchronisation der Uhren. Diese konnte nur durch 
Lichtsignale hergestellt werden, war also sehr wetterabhängig. Da der Hohe Meißner in 
Nordhessen von Göttingen und von Marburg aus bei klarem Wetter sichtbar ist, und der Feldberg 
im Taunus von Mannheim und Marburg aus erspäht werden kann, nutzte Gerling diese beiden Berge 
als Signalstationen. Dort stationierte Mitarbeiter hatten die Aufgabe, in der späten 
Nachmittagssonne mit Heliotropen Signalfolgen im Abstand von $8$ Minuten gleichzeitig an die 
jeweils benachbarten Sternwarten zu schicken. Mit der einbrechenden Nacht wurden dann 
Pulversignale gezündet. Die Signale vom Hohen Meißner und vom Feldberg wurden um $4$ Minuten 
versetzt verschickt, so dass am Frauenberg alle $4$ Minuten ein Lichtsignal aus der südwestlichen 
oder nordöstlichen Richtung ankam. Dies wurde während des angesetzten Zeitrahmens im 
Spätsommer $1837$ an jedem möglichen Nachmittag und Abend wiederholt. Tagsüber wurde die Zeit 
mit korrespondierenden Sonnenhöhenmessungen und nachts mit Sterndurchgangsmessungen festgehalten 
und so der Gang der Uhren kontrolliert.
Wetterbedingt waren natürlich nicht alle verschickten Signale beobachtbar. Von den insgesamt 
vermessenen $256$ Signalen vom Meißner und $136$ vom Feldberg bildeten $116$ korrespondierende Signale 
beider Stationen die sicherste Basis für die Synchronisation der Uhren.

Ein weiterer wichtiger Bestandteil von Gerlings Projekt war die Verringerung der Messfehler 
durch eine sehr präzise Bestimmung der Reaktionszeit der Beobachter, der so genannten 
persönlichen Gleichung. Zeitmessungen erfolgten immer durch visuelle Beobachtung eines Durchgangs. 
Der Beobachter sieht den auf eine Linie zulaufenden Stern und gibt genau beim Durchgang ein 
Zeichen, bei dem ein Mitarbeiter die Zeit an einer Uhr abliest. Solch eine Messung ist nicht 
wiederholbar, kann nicht überprüft werden. Man kann nur hoffen, den Messfehler durch 
Wiederholungen statistisch ermitteln zu können. Gerling hat dazu die beobachtenden Kollegen 
in Göttingen und Mannheim besucht und mit ihnen gleichzeitig Sterndurchgänge und Durchgänge 
eines Pendels vermessen. Dadurch bestimmte er nicht nur die Größe des Messfehlers jedes der 
Beobachter, sondern auch deren persönliche Reaktionszeit, ein Offset der Zeitmessung. Der 
Offset war beträchtlich, so dass Gerling dies in einem Brief an Gauß (Brief Nr.\ $294$, 
Schäfer $1927$) 
besonders erwähnt und damit frühere Längenbestimmungen zwischen Helgoland und Greenwich und Paris 
und Greenwich in Frage stellt!

Zum Abschluss erhielt er als Ergebnis folgende Längenunterschiede, von ihm angegeben im Zeitmaß:

\vspace*{-2ex}
\begin{tabbing}
Göttingen-Frauenberg: \hspace*{3em} \= $4$\am $36,19$\as $\pm$ $0,0152$\as\\
Frauenberg-Mannheim:  \> $1$\am $19,67$\as $\pm$ $0,0208$\as\\
Göttingen-Mannheim:	  \>	$5$\am $55,86$\as $\pm$ $0,0258$\as.
\end{tabbing}

\vspace*{-2ex}
Gerling schaffte es also immerhin, den Messfehler auf etwa $0,025$ Sekunden in der Zeitmessung und 
damit auf ca.\ $0,4$\ds in Winkeleinheiten zu begrenzen.

Mit diesem Ergebnis endet die Publikation Gerlings in den Astronomischen Nachrichten --- und die 
wirkliche Bedeutung dieser Messungen wird ihm erst im folgenden Briefkontakt mit Carl Friedrich 
Gauß klar! 
Er äußert Gauß gegenüber die deutliche Abweichung der astronomischen Längenbestimmung von der 
geodätischen. Daraufhin verweist Gauß auf seine Lotabweichungsmessung zwischen Altona und 
Göttingen und die von ihm bereits $1828$ veröffentlichten Gedanken zur erwarteten Unregelmäßigkeit 
der Erde (Brief Nr.\ $296$, Schäfer $1927$). Es ist diese Anmerkung, die Gerling die Augen vollends 
öffnet: er hat die erste Lotabweichungsmessung in der Länge durchgeführt und die gefundene
Differenz der astronomischen und geodätischen Längen stellt ein neues Ergebnis dar. Diese 
Messung hat eine Tür zu einer neuen Qualität von Messungen aufgestoßen! Er schreibt in seiner 
Antwort an Gauß (Brief Nr.\ $297$, Schäfer $1927$): 

,,\textit{Wenn ich mich hier also eines großen Irrtums und eines Mangels an Gründlichkeit in 
Benutzung 
Ihrers § \em [gemeint ist ein Abschnitt aus Gauß' Publikation von $1828$] \em selbst anklagen muß, so 
kann ich mich doch vielleicht damit trösten, daß wahrscheinlich keine ,,$\mathit{5}$ Menschen in Europa 
existieren'', welche den § in diesem Sinne beherzigt haben. \ldots Ich sehe mich deshalb zunächst 
veranlaßt, von meiner eigenen Arbeit in dieser Beziehung einen anderen und vernünftigeren Gebrauch 
zu machen, als ich anfangs nach nicht gehörig aufgeklärten Begriffen beabsichtigte; und ist 
dieses ein neues großes Verdienst, welches Sie um dieser Arbeit haben.}''

\captionsetup{justification=justified}
\begin{figure}[t]
\begin{minipage}[b]{0.40\textwidth}
\caption[normal]{Auszug aus dem Manuskript von Gerlings Veröffentlichung zur Längendifferenzmessung 
(Gerling $1839$): Es handelt sich um den Anfang des $5$.\ Kapitels. Links am Rand ist eine nicht 
maßstabsgerechte Skizze zu finden, die nicht in der Publikation enthalten ist. 
(Universitätsbibliothek Marburg, Nachlass von Christian Ludwig Gerling. 
Ms.\ $35$2, Blatt $3$v; Nachdruck mit freundlicher Genehmigung).}
\end{minipage}
\hfill
\begin{minipage}[b]{0.57\textwidth}
\includegraphics[angle=0,width=\textwidth]{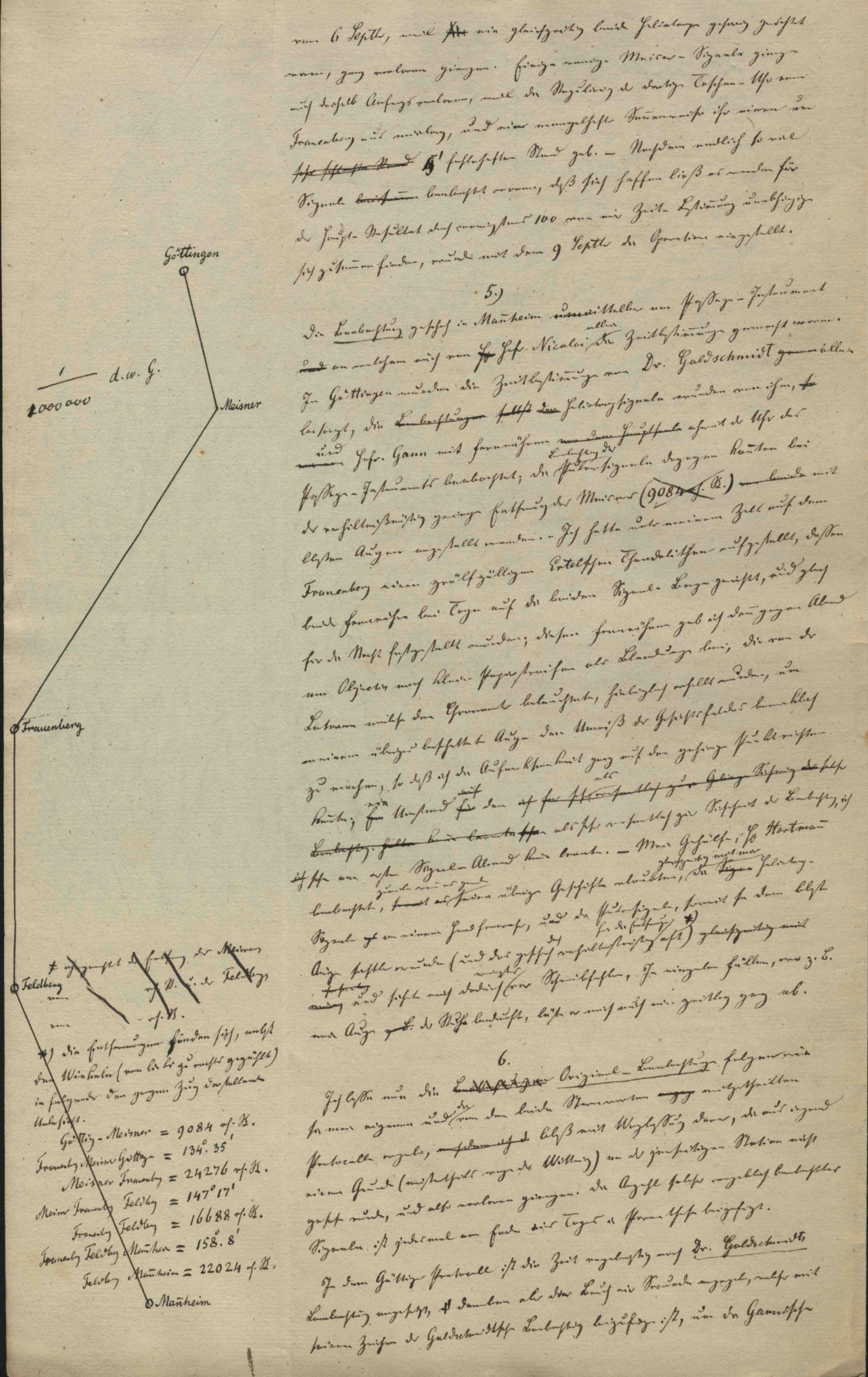}
\end{minipage}
\end{figure}

In der $1839$ erschienenen Veröffentlichung der Resultate der kurhessischen Triangulierung fand 
Gerling dann andere sehr deutliche Worte (Gerling $1839$, Seite $204$ ff) als in seiner Publikation 
von $1838$. Diesmal weist er auf den qualitativen Unterschied der astronomischen und geodätischen 
Messung hin und führt als Ergebnis die gefundenen Differenzen in den Längen auf:
\pagebreak

\addtocounter{footnote}{1}
\footnotetext[\value{footnote}]{Gerling hat diese Zahl nicht angegeben. Sie wurde der Vollständigkeit halber 
aus den anderen Daten der Tabelle errechnet.}
\addtocounter{footnote}{1}
\footnotetext[\value{footnote}]{Das gleiche gilt für diese Angabe.} 

\begin{center}
\begin{tabular}{|p{3.2cm}|p{2.23cm}|p{2.23cm}|p{2.23cm}|}
\hline
 	& astronomische Längendifferenz & geodätische\ \ \  Längendifferenz & Abweichung \\ \hline
 Göttingen-Frauenberg &	$1$\dg $09$\dm $02,85$\ds &	$1$\dg $09$\dm $19,49$\ds &	 -- $16,6$\ds\\ \hline
 Frauenberg-Mannheim  &	\hspace{1em} $19$\dm $55,05$\ds 
 \addtocounter{footnote}{-1}$^{\decimal{footnote}}$\addtocounter{footnote}{1} &  
 \hspace{1em}   $19$\dm $42,85$\ds $^{\decimal{footnote}}$
  &  $+12,2$\ds\\ \hline
 Göttingen-Mannheim	  & $1$\dg $28$\dm $57,90$\ds &	$1$\dg $29$\dm $02,32$\ds &	 -- $4,4$\ds\\ \hline
\end{tabular}
\end{center}

Die gefundenen Unterschiede zwischen astronomischen und geodätischen Längendifferenzen sind im 
Rahmen von Gerlings angegebenen Fehlern signifikant und passen gut zur Größenordnung der 
von Gauß festgestellten Breitenabweichungen in Norddeutschland.

\section{Vergleich mit späteren Messungen}
In den folgenden Jahren ergänzten Gerling und seine Mitarbeiter diese ersten Messungen der 
Längendifferenzen noch um Angaben zur Lage der $1841$ in Betrieb genommenen Sternwarte am 
Renthof, am Schlossberg in Marburg. Nach ihrer Einrichtung bestimmte Gerling zunächst durch 
eine kleine Triangulation die geodätische Position der Beobachtungssäule seiner Sternwarte 
(Gerling $1843$, $2$) wie folgt: Länge von Ferro\footnote{Zu Gerlings Zeiten wurden die 
geographischen Längen noch auf den Meridian von Ferro (El Hierro) bezogen, der 
westlichsten der  Kanarischen Inseln. Der  (damals angenommene!) Meridian von Ferro 
liegt \linebreak $17$\dg\hspace{-0.5ex}$40$\dm$\hspace{-0.5ex}00$\ds westlich von Greenwich, 
dem heutigen (astronomischen) Nullmeridian.} = $26$\dg $26$\dm $2,1$\ds\hspace{-0.5ex}, 
Breite = $50$\dg $48$\dm $46,9$\ds\hspace{-0.5ex}. 

Die astronomische Länge der Sternwarte ermittelte Ernst Wilhelm Klinkerfues, ein Student von 
Gerling und späterer Nachfolger von Gauß als Direktor der Sternwarte in Göttingen, 
aus vergleichenden Sternbedeckungen mit Bezug auf Berlin als Referenz (Gerling $1855$). 
Er fand die astronomische Länge der Sternwarte Gerlings: $18$\am $28,38$\as westlich von Berlin. 
Interessant ist, dass Klinkerfues in dieser Publikation mit folgender Anmerkung zitiert 
wird: ,,\textit{Da es aber nur sehr wenig Orte gibt, deren Länge eben so gut 
oder gar besser bestimmt ist, als die von Marburg, so erschien es mir nicht gut gethan, 
alle Beobachtungen, für den Zweck wenigstens, auf welchen ich mich beschränkte, zu 
berücksichtigen. Selbst die an Hauptsternwarten angestellten correspondierenden Beobachtungen 
habe ich nicht zugezogen, wenn, wie in zwei Fällen, die Correctionen der Tafeln durch 
Greenwicher Meridianbeobachtungen bekannt waren.}'' 
Damit wird erneut die Schwierigkeit guter Längenmessungen hervorgehoben.

Und schließlich unternimmt Richard Mauritius im Rahmen seiner Dissertation (Mauritius $1862$) 
eine astronomische Breitenbestimmung der Sternwarte Gerlings mithilfe von Messungen im 1.\ Vertikal, 
einer von Bessel und anderen eingeführten sehr fehlerarmen und somit genauen Methode zur 
Bestimmung der Polhöhe. Mauritius fand für die astronomische Breite der Sternwarte 
Gerlings: $50$\dg $48$\dm $44,09$\ds\hspace{-0.5ex}.

Die astronomische Länge berechnet er mit dem Ergebnis von Klinkerfues und einer Angabe der Länge 
von Berlin aus dem Berliner Jahrbuch. Allerdings verwendet er dazu die geodätische Position der 
Sternwarte in Berlin ($30$\dg $03$\dm $30$\ds östlich von Ferro) statt der ebenfalls angegebenen 
zeitlichen Lage östlich von Greenwich ($53$\am $35,5$\as entsprechend 
$30$\dg $03$\dm $53,375$\ds östlich von 
Ferro). Er fand also die astronomische Länge der Sternwarte Gerlings bezogen auf 
Ferro: $26$\dg $26$\dm $24,3$\ds\hspace{-0.5ex}. Somit konnte Mauritius nun für die Sternwarte in Marburg eine 
Abweichung der Längendifferenzen zu Berlin von $+22,2$\ds und eine Lotabweichung in der 
Breite von $+2,81$\ds angeben.

Die Ergebnisse der Längendifferenzmessungen sind nicht zu verwechseln mit Angaben zur 
Lotabweichung, denn es sind ja Vergleiche der Messdaten zweier entfernter Messstationen. 
Um die wahren Lotabweichungen zu erhalten, muss man offensichtlich ein möglichst dichtes Netz 
von Messstationen über das ganze Land legen. Jede Messstation muss ein Ergebnis für eine 
geodätische und eine astronomische Lagebestimmung liefern. Genau dies war ein wichtiger 
Aspekt der mitteleuropäischen Gradmessung (Baeyer $1861$), zu der Baeyer ab $1862$ viele 
europäische Länder gewinnen konnte. Im Unterschied zu Gerlings Messungen konnten die 
Wissenschaftler sich nun allerdings schon der Telegrafie bedienen, d.h.\ die sehr aufwendige 
Synchronisation der Uhren über Lichtsignale entfiel komplett. Gerlings Messung ist in 
dieser Hinsicht einmalig. Es blieb aber im $19$.\ Jahrhundert nach wie vor die Schwierigkeit 
der visuellen Beobachtung, d.h. der Eliminierung von Reaktionszeiten der Beobachter. 
Argelander hatte daher sehr deutlich angeregt, ,,\textit{sämtliche Polhöhen und 
Längenbestimmungen über das ganze zu untersuchende Areal der Gradmessung von denselben 
Beobachtern, etwa $\mathit{4}$ an der Zahl, und mit völlig gleichen Instrumenten ausführen zu lassen}'', 
was leider in der Praxis nicht ,,\textit{in aller Strenge hat durchgeführt werden können}'' 
(Hilfiker $1885$). Albrecht, Bruns und Hilfiker beschäftigten sich mit der Berechnung 
des Längennetzes in Europa und veröffentlichten die Ergebnisse zu den astronomischen 
Längenbestimmungen der Sternwarten (Hilfiker $1885$). Aber das Netz dieser Längenmessungen 
war immer noch nicht ausreichend dicht. Erst im $20$.\ Jahrhundert gelang es mit modernen 
Gravimetern und Zenitkameras sowohl gravimetrisch als auch astronomisch ein wesentlich 
dichteres Netz von Messpunkten zu realisieren und so Karten von Lotabweichungen zu erstellen. 

Die folgende Tabelle $1$ enthält eine Aufstellung von Gerlings Messungen und einen Vergleich 
mit den Längenbestimmungen der mitteleuropäischen Gradmessung sowie mit modernen Daten.

Die geodätischen Koordinaten auf dem heutigen globalen Referenzellipsoid GRS$80$ beziehen sich 
auf das Europäische Terrestrische Referenzsystem $1989$ (ETRS$89$) und wurden vom Hessischen
Landesamt für Bodenmanagement (HLBG) vom Landesamt für Geoinformation und Landentwicklung 
Niedersachsen (LGLN) sowie vom Landesamt für Geoinformation und Landeentwicklung 
Baden-Württemberg (LGL) zur Verfügung gestellt. Die Lotabweichungen auf dem GRS$80$ wurden vom 
Bundesamt für Kartographie und Geodäsie (BKG) aus dem aktuellen Geoidmodell für Deutschland 
unter zusätzlicher Berücksichtigung der lokalen Massenverteilungen im Umfeld der betreffenden 
Stationen ermittelt. Die astronomische Länge der Sternwarte Marburg berechnete Mauritius 
aus der von Klinkerfues ermittelten astronomischen Längendifferenz zu Berlin, verwendete 
dazu jedoch die Angabe der geodätischen Lage im Berliner Jahrbuch. Zur Einschätzung der 
Qualität der Marburger Messungen sind nur die jeweils grau schattierten Angaben zu den 
astronomischen Breiten und den astronomischen Längendifferenzen relevant. Zum einen zeigt 
die Tabelle den Fortschritt in der Erhebung der Daten. Die Abweichungen der einzelnen 
Positionsangaben liegen in der Regel über den angegebenen Messfehlern --- was für darin 
versteckte systematische Fehler spricht. 

Die astronomische Längendifferenz zwischen Göttingen und Mannheim wurde von Gerling sehr 
präzise ermittelt, die Abweichung zu den heutigen Daten beträgt $0,75$\ds\hspace{-0.5ex}! Eine deutliche 
Abweichung von $10,07$\ds zeigt sich jedoch bei der Längendifferenz zum Frauenberg hin. 
Gerling errechnete die gesamte Differenz zwischen Göttingen und Mannheim aus der Summe 
der Werte von Göttingen zum Frauenberg und der vom Frauenberg nach Mannheim. Daher 
spricht eine zu große Differenz jeweils zum Frauenberg hin deutlich für einen Offset 
in der Bestimmung der lokalen Sternzeit bzw.\ mittleren Zeit am Frauenberg. Gerling selber 
erwähnte zu genau diesem Punkt, dass der eigens für die Zeitmessungen errichtete Steinpfosten 
sich nicht als stabil genug erwies und die dort aus Sterndurchgängen ermittelten Zeiten 
keine Verbesserungen zu seinen aus korrespondierenden Sonnenhöhen bestimmten Zeiten ergaben. 
Die Sonnenhöhen hat Gerling mit einem Steinheilschen Prismenkreis und einem künstlichen 
Horizont gemessen. Mit der daraus berechneten mittleren Zeit kontrollierte er den Lauf 
seiner Präzisionsuhr. So konnte er einen Sprung in der täglichen Gangdifferenz im Bereich 
von $0,2$ Sekunden nach einem stürmischen und kalten Tag nachweisen; aber vermutlich ist der 
Fehler letztendlich doch in der mit dem Prismenkreis bestimmten Zeit zu suchen. Solch 
präzise Zeitmessungen sind mit standfesten und größeren Geräten sehr viel besser durchzuführen 
als mit einem relativ kleinen Prismenkreis. So sind die Zeitbestimmungen in Göttingen und 
Mannheim sehr gut und eine ganz wesentliche Grundlage für Gerlings Ergebnis!

Gerlings astronomische Längendifferenzmessung zwischen Göttingen und Mannheim ist aber 
beispiellos perfekt und zu bewundern; der große Aufwand zur Synchronisation der Uhren hat 
sich gelohnt. Gerling konnte entgegen der zurückhaltenden Meinung von Gauß zeigen, dass 
es mit den Methoden in der ersten Hälfte des $19$.\ Jahrhunderts wirklich möglich war, den 
Lotabweichungen auf die Spur zu kommen. In diesem Sinne gebührt ihm neben C.F. Gauß ein 
respektabler Platz in der Geschichte der Erforschung der Lotabweichungen unserer Erde.

\newpage

\begin{sidewaystable}
\footnotesize
\renewcommand{\arraystretch}{1.3}
\begin{tabularx}{\textwidth}{|C{1.33}|C{0.945}|C{0.945}|C{0.945}|C{0.945}|C{0.945}|C{0.945}|}
\hline
 & Göttingen, Sternwarte & Frauenberg, Steinpfeiler & Marburg, Sternwarte & Mannheim, Sternwarte &
 Längendifferenz Göttingen-Mannheim & Längendifferenz Göttingen-Frauenberg \\ \hline
geodätische Breite, Walbeck & $51$\dg $3$1\dm $47,850$\ds & $50$\dg $45$\dm $27,751$\ds 
& $50$\dg $48$\dm $46,884$\ds & $49$\dg $29$\dm $14,681$\ds & & \\ 
astronomische Breite, $1862$ & & &\cellcolor[gray]{0.9} $50$\dg $48$\dm $44,09$\ds 
(Mauritius $1862$) & & &\\
geodätische Breite, GRS$80$ & $51$\dg $31$\dm $42,943$\ds & $50$\dg $4$5\dm $23,494$\ds & 
$50$\dg $48$\dm $42,583$\ds & $49$\dg $29$\dm $11,385$\ds & & \\
Lotabweichung $\xi$,\ \ \ \  BKG $2012$ & $+4,813$\ds & $-0,168$\ds &
$+0,765$\ds & $-0,013$\ds & & \\ \hline astronomische Breite, heute & $51$\dg
$31$\dm $47,756$\ds & $50$\dg $45$\dm $23,326$\ds & \cellcolor[gray]{0.9} $50$\dg $48$\dm $43,348$\ds& $49$\dg $29$\dm $11,372$\ds & & \\ \hline
geodätische Länge (Ferro), Walbeck & $27$\dg $36$\dm $28,200$\ds & $26$\dg $27$\dm $08,712$\ds 
& $26$\dg $26$\dm $02,100$\ds & $26$\dg $07$\dm $27,712$\ds & $1$\dg $29$\dm
$02,32$\ds & $1$\dg $09$\dm $19,49$\ds \\ 
astronomische Länge, $1839$/$1862$ & $27$\dg $36$\dm $26,40$\ds (Ref.\ \ \ Mannheim) 
& $26$\dg $27$\dm $23,55$\ds (Ref.\ \ \ Mannheim)
& $26$\dg $26$\dm $24,3$\ds ($1862$, Ref.\ Berlin)& $26$\dg $07$\dm $28,5$\ds (Nicolai/Wurm) 
& \cellcolor[gray]{0.9} $1$\dg $28$\dm $57,90$\ds & \cellcolor[gray]{0.9} $1$\dg $09$\dm $02,85$\ds \\
astronomische Länge, $1885$& $27$\dg $36$\dm $35,944$\ds & & & $26$\dg $07$\dm $38,689$\ds 
& $1$\dg $28$\dm $58,255$\ds & \\
geodätische Länge (Ferro), GRS$80$ & $27$\dg $36$\dm $34,022$\ds & $26$\dg $27$\dm $15,677$\ds 
& $26$\dg $26$\dm $09,080$\ds & $26$\dg $07$\dm $34,912$\ds & $1$\dg $28$\dm $59,110$\ds & \\
Lotabweichung $\eta/\cos\varphi$, BKG $2012$ & $-0,775$\ds & $+4,651$\ds & $+6,566$\ds 
& $+1,184$\ds & & \\ \hline
astronomische Länge (Ferro), heute & $27$\dg $36$\dm $33,247$\ds & $26$\dg $27$\dm $20,328$\ds
& $26$\dg $26$\dm $15,646$\ds & $26$\dg $07$\dm $36,096$\ds 
& \cellcolor[gray]{0.9} $1$\dg $28$\dm $57,151$\ds 
& \cellcolor[gray]{0.9} $1$\dg $09$\dm $12,919$\ds \\ \hline
\end{tabularx}
\caption{Zusammenstellung verschiedener Breiten- und Längenmessergebnisse}
\renewcommand{\arraystretch}{1.0}
\end{sidewaystable}

\clearpage
{ 
\footnotesize
\baselineskip9.5pt
\parindent0em
\textbf{Danksagung:}
Am $9$.\ September $2012$ jährte sich das Ende von Gerlings Längendifferenzmessungen am Frauenberg 
zum $175$.\ Male. Herr Hans-Jürgen Will, ehemaliger Mitarbeiter im Hessischen Landesamt für 
Bodenmanagement und Geoinformation Wiesbaden (HLBG), regte aus diesem Anlass am Tag des 
offenen Denkmals 2012 eine öffentliche Präsentation von Gerlings astronomisch-geodätischen 
Aktivitäten an. Dies war auch der Anstoß zur Aufbereitung der historischen Unterlagen für 
diese Publikation, dafür sei ihm herzlichst gedankt. Herr Bernhard Heckmann (HLBG) trug durch 
fundierte Diskussionsbeiträge sowie durch die Beschaffung der aktuellen Positions- und 
Lotabweichungsdaten wesentlich zur Klärung und Einordnung von Gerlings Messungen 
bei --- auch dafür herzlichen Dank. Und nicht zuletzt sei auch dem Bundesamt für Kartographie 
und Geodäsie (BKG) -- Außenstelle Leipzig --- Dank für die Ermittlung und Bereitstellung 
aktueller Lotabweichungsangaben ausgesprochen.

\vspace{-1.0em}
\noindent
\section*{Literatur:}
\begin{list}{}{%
\setlength{\topsep}{0pt}%
\setlength{\leftmargin}{3.5mm}%
\setlength{\listparindent}{-3.5mm}%
\setlength{\itemindent}{-3.5mm}%
\setlength{\parsep}{0.3\parskip}%
}%
\item[]
Alder, Ken ($2005$): ,,\textit{Das Maß der Welt – Die Suche nach dem Urmeter}'',
	 Goldmann Verlag, München, $1$.\ Aufl.
	
Baeyer, Johan Jacob ($1861$): ,,\textit{Über die Größe und Figur der Erde}'', 
	Verlag von Georg Reimer, Berlin.
	
Baeyer, Johann Jacob ($1864$): ,,\textit{General-Bericht über die mitteleuropäische 
	Gradmessung pro $1863$}'', Verlag von Georg Reimer, Berlin.

Baeyer, Johann Jacob ($1866$): ,,\textit{General-Bericht über die mitteleuropäische Gradmessung für das 
	Jahr $1865$}'', Verlag von Georg Reimer, Berlin.

Bessel, Friedrich ($1837$): ,,\textit{Bestimmung der Axen des elliptischen Rotationssphäroids, welches 
	den vorhandenen Messungen von Meridianbögen der Erde am meisten entspricht}'', Astron.\ 
	Nachr.\ $14$, $333$.
	
Bessel, Friedrich ($1841$): ,,\textit{Über einen Fehler in der Berechnung der französischen Gradmessung 
	und seinen Einfluss auf die Bestimmung der Figur der Erde}'', 
	Astron.\ Nachr.\ $19$, $97$.

Gauß, Carl Friedrich ($1828$): ,,\textit{Bestimmung des Breitenunterschiedes zwischen den 
	Sternwarten von Göttingen und Altona}'', Vandenhoeck und Ruprecht, Göttingen. 
	Nachdruck in: C.F. Gauß Werke, $9$.\ Bd., B.G.\ Teubner, Leipzig $1903$.
	
Gerardy, Theo (Hrsg.) ($1964$); ,,\textit{Christian Ludwig Gerling an Carl Friedrich Gauss. Sechzig
	bisher unveröffentlichte Briefe}'', (=Arbeiten aus der Niedersächsichen Staats- und 
	Universitätsbibliothek Band $5$), Vandenhoeck und Ruprecht, Göttingen ($1964$).
	
Gerling, Christian Ludwig ($1838$): ,,\textit{Die Längenunterschiede zwischen Göttingen (Altona), 
	Marburg und Mannheim bestimmt durch Lichtsignale}'', 
	Astron.\ Nachr.\ $15$ ($351$\&$352$), $249$.
	
Gerling, Christian Ludwig ($1839$): ,,\textit{Beiträge zur Geographie Kurhessens und der umliegenden 
	Gemeinden}'', Johann Christian Krieger’s Verlagsbuchhandlung, Cassel.
	
Gerling, Christian Ludwig ($1843$): ,,\textit{Die Ausgleichsrechnungen der practischen Geometrie oder 
	die Methode der kleinsten Fehlerquadrate in der Anwendung für geodätische Aufgaben}'', 
	F. und A. Perthes, Hamburg und Gotha.
	
Gerling, Christian Ludwig ($1843$): ,,\textit{Geodätische Festlegung des Dörnberger-Hof-Thurms zu 
	Marburg}'', Astron.\ Nachr.\ $20$ ($458$), $25$.
	
Gerling, Christian Ludwig ($1855$): ,,\textit{Lage von Marburg aus Sternbedeckungen}'', 
	Astron.\ Nachr.\ $40$ ($954$), $293$.
	
Heckmann, Bernhard ($2012$): ,,\textit{Die Gerling’sche Haupttriangulation von Kurhessen – 
	neuere Erkenntnisse und Wiederentdeckungen}'', DVW-Mitteilungen Hessen/Thüringen $1$/$2012$, $2$.
	
Hilfiker, Jakob ($1885$): ,,\textit{Ausgleichung des Längennetzes der Europäischen Gradmessung}'', 
	Astron.\ Nachr.\, $112$, $145$.
	
Kater, Henry ($1818$): ,,An account of experiments for determining the length of 
	the pendulum vibrating seconds in the latitude of London''. 
	Phil.\ Trans.\ R.\ Soc.\ (London) $104$ ($33$): p.\ $109$.
	
Laplace, Pierre Simon ($1799$), ,,\textit{Traité de Méchanique Céleste}'', Bd.\ $2$, VII, Paris.

Madelung, Otfried ($1996$): ,,\textit{Das mathematisch-physikalische Institut der 
	Universität Marburg $\mathit{1800}$ bis $\mathit{1920}$}'', Selbstverlag des Fachbereich Physik der 
	Philipps-Universität Marburg.
	
Mauritius, Richard ($1862$): ,,\textit{Bestimmung der Polhöhe von Marburg}'', Dissertation, Marburg.

Nicolai, Friedrich Bernhard ($1825$): ,,\textit{Resultate von Pulver-Signalen zu 
	geographischen Längen-Bestimmungen in der Gegend von Mannheim, im 
	Sommer $\mathit{1824}$ angestellt}'', Astronomisches Jahrbuch für das Jahr $1828$, 
	Bd.\ $53$, Seite $127$, Hrsg. J.E.\ Bode, Berlin.
	
Reinhertz, Carl ($1901$): ,,\textit{Christian Ludwig Gerling’s geodätische Thätigkeit}'', 
	Zeitschr.\ f.\ Vermessungswesen $30$.
	
Schäfer, Clemens ($1927$): ,,\textit{Briefwechsel zwischen Carl Friedrich Gauß und Christian 
	Ludwig Gerling}'', Otto Elsner Verlagsgesellschaft M.B.H., Berlin.
	
Schrimpf, Andreas; Lipphardt, Jörg; Heckmann, Bernhard (2010): ,,\textit{Wiederentdeckungen an 
	der alten Gerling-Sternwarte in Marburg}'', DVW-Mitteilungen Hessen/Thüringen $2$/$2010$, $27$.
	
Torge, Wolfgang ($2003$): ,,\textit{Geodäsie}'', Walter de Gruyter GmbH \& Co KG, Berlin, $2$. Auflage.

Torge, Wolfgang ($2009$): ,,\textit{Die Geschichte der Geodäsie in Deutschland}'', 
	Walter de Gruyter GmbH \& Co KG, Berlin, $2$. Auflage.
	
Witmann, A., Nehrkamp, H.-H., Quaiser, R., Kompart, H.\ ($2010$), ,,\textit{Der ,Gaußsche Meridian'
	Göttingen-Altona}'', Mitt.\ Gauß-Ges.\ $47$, S.\ $63-81$.
\end{list}
}

\vspace*{3ex}
\noindent
Anschrift des Verfassers:

\noindent
Priv.\ Doz.\ Dr.\ Andreas Schrimpf\\
Fachbereich Physik der Philipps-Universität Marburg\\
Renthof $5$\\
D-$35032$ Marburg\\
e-mail: andreas-schrimpf@physik.uni-marburg.de

\end{document}